# Exceptional points in parity-time symmetric plasmonic Huygens' metasurfaces

ANDREW BUTLER[1] AND CHRISTOS ARGYROPOULOS[1,2,*]

[1]*Department of Electrical and Computer Engineering, University of Nebraska-Lincoln, Lincoln, Nebraska 68588, USA*
[2]*Department of Electrical Engineering, The Pennsylvania State University, University Park, PA 16803, USA*
*\*christos.argyropoulos@unl.edu*

**Abstract:** Parity-Time (PT) symmetric optical structures exhibit several unique and interesting characteristics with the most popular being exceptional points. While the emerging concept of PT-symmetry has been extensively investigated in bulky photonic designs, its exotic functionalities in nanophotonic non-Hermitian plasmonic systems still remain relatively unexplored. Towards this goal, in this work we analyze the unusual properties of a plasmonic Huygens' metasurface composed of an array of active metal-dielectric core-shell nanoparticles. By calculating the reflection and transmission coefficients of the metasurface under various levels of gain, we demonstrate the existence of reflectionless transmission when an exceptional point is formed. The proposed new active metasurface design has subwavelength thickness and can be used to realize ultracompact perfect transmission optical filters.

## 1. Introduction

Parity-Time (PT) symmetric optical structures have been at the forefront of photonic research thanks to their unique and interesting functionalities [1]. In most cases, PT-symmetry can be realized in optics by using a balanced spatial arrangement of gain and loss materials [2]. Exceptional point (EP) singularities have been observed in various PT-symmetric photonic systems when their eigenvalues converge and coalesce in a singular point in the parametric space, causing unidirectional reflectionless transmission and other distinctive phenomena [3–6]. Even more interesting optical responses have been realized by encircling the EP singularity resulting in polarization conversion and mode transitions in chiral and multimode systems, respectively [7–10]. These unique characteristics are appealing for a plethora of emerging photonic applications including optical filtering [11].

While EPs have been extensively investigated in bulky photonic designs [12–15], their exotic functionalities in nanophotonic non-Hermitian plasmonic systems still remain relatively unexplored. Some recent notable efforts towards this goal are the investigation of active and passive metallic gratings [16,17], plasmonic nanostructures [18–21], graphene [22,23], and epsilon-near-zero materials [24–28]. An alternative promising class of nanophotonic non-Hermitian scatterers are core-shell plasmonic nanoparticles loaded with active (gain) materials. The active material in such structures can enhance scattering and compensate the inherent optical losses of metallic (plasmonic) cores or shells [29–31]. Recently, EPs have theoretically been demonstrated in core-shell nanoparticles distributed in a honeycomb pattern, though the requirement of varying the gain between multiple particles makes an experimental verification of this concept very challenging [32]. EPs have also been theoretically demonstrated in coupled chains of polarizable nanoparticles [33]. Here, we utilize an alternative approach by using the unique properties of an active PT-symmetric non-Hermitian Huygens' plasmonic metasurface design to realize the formation of EPs.

The emerging field of Huygens' metasurfaces has been in the research spotlight during recent years from microwave to optical frequencies [34,35]. Huygens' sources can be achieved by an array of resonators with orthogonal and equal strength electric and magnetic dipole responses. The induced symmetric electric dipole and anti-symmetric magnetic dipole interfere and eliminate radiation in one direction leading to relatively high transmission. However, the inherent losses of metals or high index dielectrics preclude its design in the visible spectrum and the vast majority of Huygens' metasurfaces have been proposed in near-infrared and lower microwave frequencies [34,35].

In the current work, we propose a new design of an active PT-symmetric Huygens' plasmonic metasurface operating in the visible regime. It is composed of a two-dimensional (2D) planar array of active metal-dielectric core-shell spherical nanoparticles. We demonstrate the existence of EPs in this novel PT-symmetric nanophotonic system that, interestingly, has subwavelength thickness on the contrary to alternative widely investigated relatively bulky photonic PT-symmetric configuration. Using the well-established Mie theory, we calculate the electric and magnetic dipole response of a single core-shell passive nanoparticle, proving the existence of a Huygens' source point. Then, by extending the coupled dipole approximation to active systems, we calculate the reflection and transmission coefficients of the resulting ultrathin metasurface with and without gain. Next, we explicitly demonstrate the EP formation dynamics when a small gain coefficient is introduced in the core of each spherical nanoparticle. We further analyze the EP singularity by demonstrating the eigenvalue switching behavior when encircling the EP in the complex frequency space. The proposed non-Hermitian PT-symmetric ultrathin metasurface design is expected to have applications in compact perfect transmission optical filtering, since its thickness is extremely smaller compared to conventional multilayer filters based on low loss dielectric materials [36–39].

## 2. Plasmonic Huygens' metasurface design

We begin our investigation by considering the spherical core-shell plasmonic nanoparticle shown in Fig. 1(a) with shell and core radii $r_s = 35nm$ and $r_c = 24nm$, respectively. The shell is made of silver [40] and the core has a dielectric constant of $\varepsilon_c = 4$. The particle is suspended in a host medium with $\varepsilon_h = 2.2$. Similar types of composite nanoparticles have been experimentally realized in previous works, such as gold coated colloidal silica ($SiO_2$) [41] and $SiO_2$-platinum (Pt)-$SiO_2$ core-shell-shell nanoparticles [42]. The electric and magnetic dipolar ($\alpha_{ed}$ and $\alpha_{md}$, respectively) and quadrupolar ($\alpha_{eq}$ and $\alpha_{mq}$, respectively) polarizabilities of the composite nanoparticle as a function of wavelength are computed by the formulas [43]:

$$\alpha_{ed} = -j\frac{6\pi}{k^3}a_1, \qquad (1)$$

$$\alpha_{md} = -j\frac{6\pi}{k^3}b_1, \qquad (2)$$

$$\alpha_{eq} = -j\frac{120\pi}{k^5}a_2, \qquad (3)$$

$$\alpha_{mq} = -j\frac{120\pi}{k^5}b_2, \qquad (4)$$

where $k$ is the wavenumber in the host medium, and [$a_1$, $a_2$, $b_1$, $b_2$] are the relevant scattering coefficients calculated by Mie theory [44,45]. The computed polarizabilities are plotted as a function of wavelength in Fig. 1(b), where their values have been normalized to the volume $V$ of the spherical composite nanoparticle. The quadrupolar response of the nanoparticle is negligible in the under study visible wavelength range, as clearly depicted in Fig. 1(b). Interestingly, the magnitudes of the electric and magnetic dipole polarizabilities are equal but not zero near λ=410nm (yellow star point in Fig. 1(b)). The coincidence in the electric and magnetic response of the core-shell nanoparticle is a typical feature of a Huygens' dipole source [34].

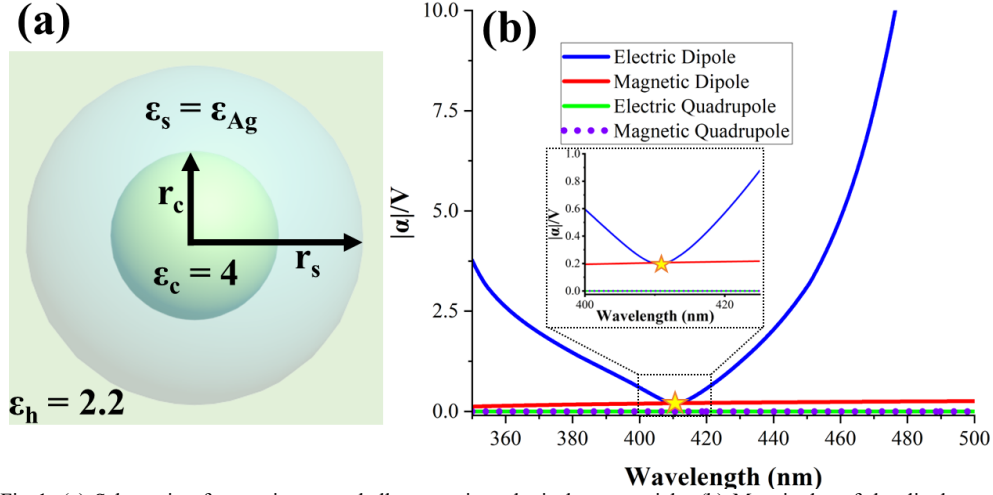

Fig 1. (a) Schematic of a passive core-shell composite spherical nanoparticle. (b) Magnitudes of the dipolar and quadrupolar polarizabilities normalized to the volume of the nanoparticle as a function of wavelength. The inset demonstrates a zoomed plot at the Huygens' point.

Next, we consider the ultrathin plasmonic metasurface illustrated in Fig. 2, composed of a 2D planar array of periodic spherical core-shell nanoparticles (with dimensions similar to Fig. 1(a)), where the periodicity between each nanoparticle is $d$ and the total metasurface thickness is $t = 2r_s = 70nm$. Again, the nanoparticles are embedded in a host material with dielectric $\varepsilon_h = 2.2$, i.e., a more realistic scenario than being suspended in air. Now, we also introduce a positive imaginary part to the core permittivity and its dielectric constant becomes complex: $\varepsilon_c = \varepsilon_r + j\varepsilon_i$, where $\varepsilon_r = 4$ is the real part and $\varepsilon_i$ is the imaginary part responsible for the material gain. Again, the dielectric constant of the plasmonic outer shell $\varepsilon_s$ is passive and taken to be silver [40]. Gain materials can be loaded into core-shell nanoparticles [29,46,47] and the nanoparticles can be assembled into an array and embedded into host matrix materials forming designs similar to the current metasurface [48–51]. Organic dye molecules are typically used to provide gain. These molecules can be introduced as dopants during the nanoparticle synthesis process [47]. Thus, the fabrication of the proposed metasurface system is experimentally feasible. Since each composite spherical nanoparticle has a Huygens' dipole response, as explained in the previous paragraph, it is expected that a planar array composed of these nanoparticles will form a Huygens' metasurface [35], which will be demonstrated in the next section.

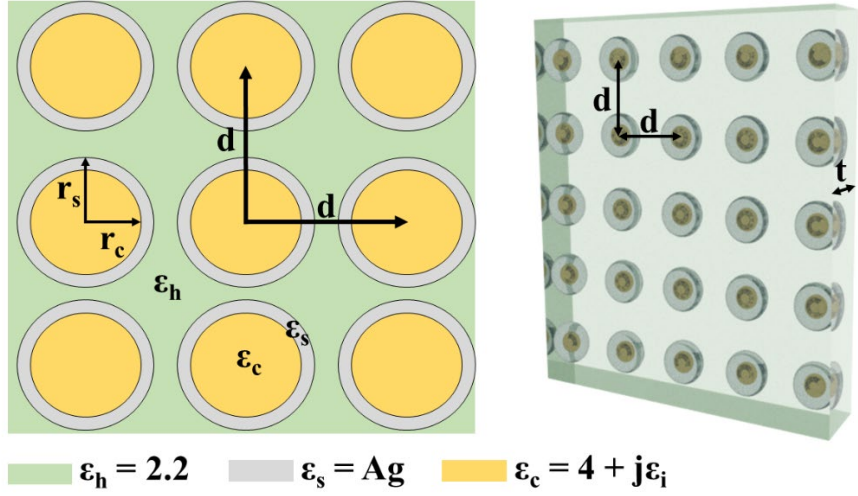

Fig 2. Schematic of the PT-symmetric plasmonic metasurface design composed of a 2D planar array of active core-shell composite spherical nanoparticles.

## 3. Exceptional point results

The response of the plasmonic metasurface presented in the previous section can be analyzed in the passive and active operation regimes through computing its reflection and transmission coefficients. These are calculated by the individual response of each nanoparticle combined with their collective response computed using the coupled dipole approximation [43]. Hence, by taking into account the contributions from both electric and magnetic dipoles of the individual scatterer, which are non-negligible as depicted before in Fig. 1(b), the reflection ($\Gamma$) and transmission ($t$) coefficients are calculated as:

$$\Gamma = \frac{-\eta_0}{2Z_{symm} + \eta_0} + \frac{Z_{asymm}}{Z_{asymm} + 2\eta_0}, \quad (5)$$

$$t = \frac{2Z_{symm}}{2Z_{symm} + \eta_0} - \frac{Z_{asymm}}{Z_{asymm} + 2\eta_0}, \quad (6)$$

where $\eta_0$ is the impedance of the metasurface's surrounding free space, $Z_{symm}$ is the effective symmetric impedance of the metasurface due to electric dipole contributions, and $Z_{asymm}$ is the effective asymmetric impedance of the metasurface due to magnetic dipole contributions [43]. The impedances $Z_{symm}$ and $Z_{asymm}$ are computed by using the formulas:

$$Z_{symm} = Z_e^d, \quad (7)$$

$$Z_{asymm} = Z_m^d, \quad (8)$$

where $Z_e^d, Z_m^d$ are the impedances due to the electric and magnetic dipoles, respectively. The formulas of these impedances are presented in [43,52] and they are computed by the polarizabilities given by Eqs. (1)-(2) and geometrical dimensions of the nanoparticle array along with the operation wavelength. The electric and magnetic quadrupole impedances are neglected in these calculations since the nanoparticle's quadrupolar responses demonstrated in Fig. 1(b) are negligible compared to the dipolar responses. The quadrupolar responses remain similarly lower compared to the dipolar responses even when gain is introduced in the nanoparticle. As mentioned before, a detailed description on how to compute these impedances can be found in previous works relevant to passive dielectric metasurfaces [43,52]. Here, we modify this existing methodology, essentially extending it to active plasmonic metasurface systems.

Hence, the reflectance and transmittance of the currently presented PT-symmetric metasurface for different values of gain can be calculated as $R=|\Gamma|^2$ and $T=|t|^2$, respectively, where $\Gamma$ and $t$ coefficients are computed by Eqs. (5)-(6). The parameters of the metasurface are identical to the Huygens' nanoparticle used before: $r_c = 24nm$, $r_s = 35nm$, but now the array period is $d = 80nm$. The dielectric constant of the host material is taken as a constant $\varepsilon_h = 2.2$, similar to the nanoparticle case. The computed reflectance and transmittance for the passive metasurface when $\varepsilon_i = 0$ (i.e., no gain) is presented in Fig. 3(a). Note that high transmission through the metasurface is obtained at the same wavelength as the Huygens' dipole source demonstrated before in Fig. 1(b) for a single nanoparticle scatterer. In this case, the reflectance and transmittance do not add up to unity since there is always absorption in the system due to the inherent Ohmic losses of the silver nanoshell, while the reflectance value is low but not exactly equal to zero. However, this response is representative to a new passive and lossy (non-Hermitian) Huygens' plasmonic metasurface design [35], where the Huygens' point is remarkably achieved deep in the visible range. The losses in the visible due to the plasmonic material can be fully compensated by introducing dielectric gain materials in the nanoparticle core. Hence, when the gain is increased to $\varepsilon_i = 0.115$, the transmittance through the now active Huygens' plasmonic metasurface system increases to unity while the reflectance drops to exactly zero. At this critical point, the proposed metasurface exhibits perfect reflectionless transmission, an interesting functionality that corresponds to the EP formation in non-Hermitian PT-symmetric systems, as it will be shown in the next paragraph. Interestingly, the formed EP also corresponds to the Huygens' dipole response of each single nanoparticle scatterer, highlighting the strong connection between these two emerging optical concepts. In addition, compared to a previous work [29] which experimentally demonstrated a very high positive imaginary part of permittivity of approximately $\varepsilon_i = 0.5$ in a composite nanoparticle, the gain needed to achieve the EP in the current metasurface is much lower, i.e., it has realistic values.

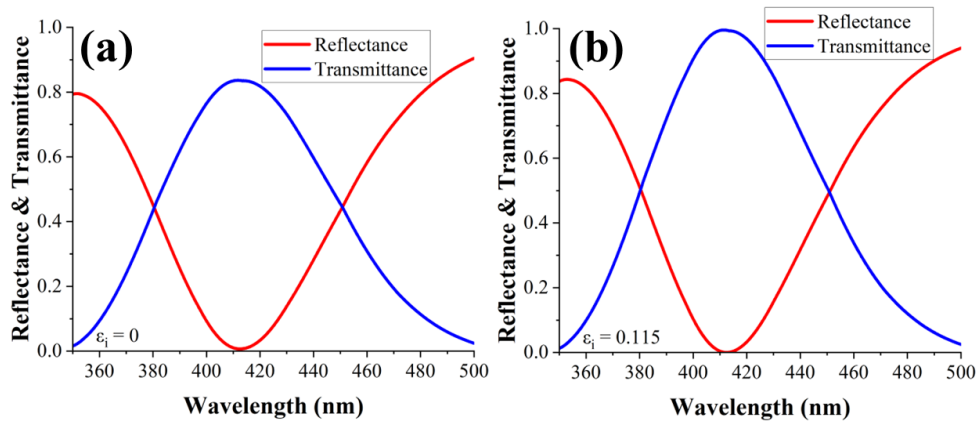

Fig 3: Reflectance and transmittance versus wavelength for the (a) passive (no gain/$\varepsilon_i = 0$) and (b) active ($\varepsilon_i = 0.115$) metasurface design, where the later corresponds to EP formation.

The EP formation dynamics of the active metasurface can be further analyzed by calculating the transfer matrix M of the system. The reflection and transmission coefficients are used to calculate the M matrix following the formula [53]:

$$M = \begin{bmatrix} M_{11} & M_{12} \\ M_{21} & M_{22} \end{bmatrix} = \begin{bmatrix} t - \frac{\Gamma^2}{t} & \frac{\Gamma}{t} \\ -\frac{\Gamma}{t} & \frac{1}{t} \end{bmatrix}. \qquad (9)$$

To unambiguously prove that the metasurface response under low gain values indeed leads to an EP formation, we calculate the eigenvalues of the M-matrix as a function of either the wavelength or the gain coefficient by using the formula [54]:

$$k_{1,2} = \frac{M_{11} + M_{22}}{2} \pm \sqrt{\left(\frac{M_{11} + M_{22}}{2}\right)^2 - 1}. \qquad (10)$$

The eigenvectors ($\phi_1$, $\phi_2$) of the M-matrix are also computed by [54]:

$$\phi_{1,2} = \begin{pmatrix} \frac{k_{1,2} - M_{22}}{M_{21}} \\ 1 \end{pmatrix}. \qquad (11)$$

The computed eigenvalues ($k_1$, $k_2$) of the system at the EP and slightly off its formation are computed on the complex frequency plane and plotted as a function of wavelength and gain in Fig. 4. In particular, the imaginary part of both eigenvalues diminishes to zero while the real part becomes equal to unity only when $\varepsilon_i = 0.115$ and $\lambda = 411nm$, as clearly demonstrated in Figs. 4(a) and 4(b), respectively. This response consists evident proof that both eigenvalues converge and coalesce to an EP singularity shown by the yellow star in Figs. 4(a) and 4(b). This EP lies on the right side of the unit circle (black dotted line) where the real part of both complex eigenvalues is unity and imaginary part is zero. On the contrary, when the gain or wavelength are slightly detuned from the EP fixed values of $\varepsilon_i = 0.115$ and $\lambda = 411nm$ to $\varepsilon_i = 0.05$ and $\lambda = 415nm$, the two eigenvalues of the transfer matrix no longer coalesce to the EP, as it is clearly depicted in Figs. 4(c) and 4(d), respectively.

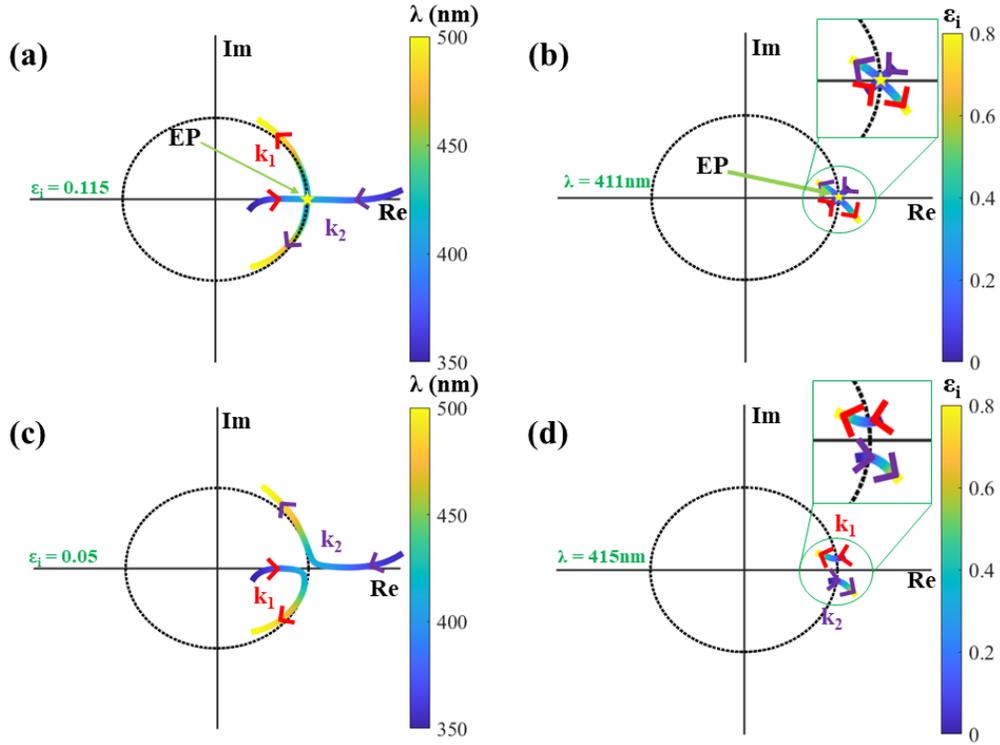

Fig 4. Transfer matrix eigenvalues near and exactly at the EP as a function of (a) wavelength and (b) gain coefficient plotted on the complex frequency plane. When the (c) wavelength or (d) gain is slightly off the EP formation, the eigenvalues no longer coalesce to this singular point.

The EP can alternatively be demonstrated by looking at the magnitudes of the eigenvalues and eigenvectors of the M-matrix when computed by using Eqs. (10) and (11), respectively. These results are presented in Fig. 5, where the calculated magnitudes of the eigenvalues and eigenvectors as a function of wavelength are plotted. Interestingly, the magnitudes of the eigenvalues and eigenvectors converge at the EP wavelength $\lambda = 411nm$ and then bifurcate (solid lines in Figs. 5(a) and 5(b), respectively) only when the gain is fixed to $\varepsilon_i = 0.115$. Notably, slightly off the gain value needed to achieve the EP formation ($\varepsilon_i = 0.05$), the eigenvalues never fully merge (dotted lines in Fig. 5(a)). This is another direct proof of the EP formation dynamics present at the current PT-symmetric Huygens' plasmonic metasurface which can be used as a perfect transmission planar filter for visible light with the practical advantage of subwavelength thickness. Additional discussion on the evolution of the eigenvalues over an extended wavelength range and with increased gain can be found in the supplementary information [55].

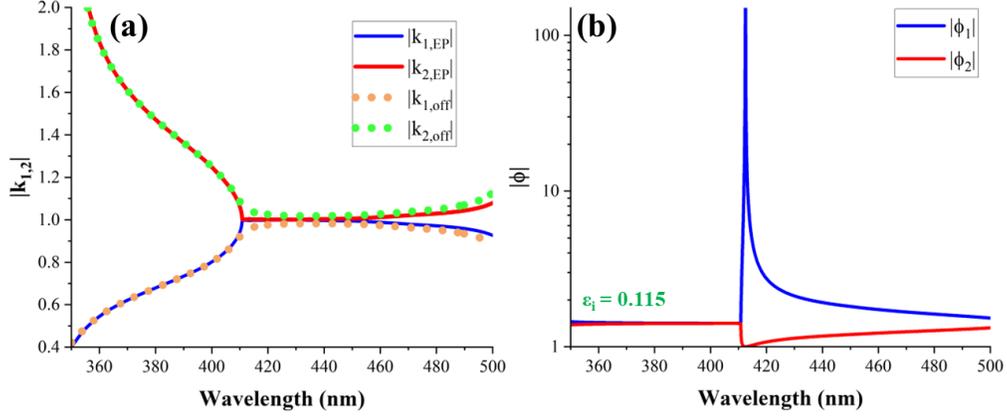

Fig 5. (a) Magnitude of the eigenvalues as a function of wavelength for a fixed gain value of $\varepsilon_i = 0.115$ (solid lines) corresponding to the EP and for $\varepsilon_i = 0.05$ (dotted lines), i.e., slightly off the EP formation. (b) Magnitude of the eigenvectors as a function of wavelength for a fixed gain value of $\varepsilon_i = 0.115$ corresponding to the EP formation.

The topological effect of encircling an EP is one of the most remarkable properties in the physics of exceptional point formation dynamics [7–10]. Hence, we demonstrate in Fig. 6 that the EP encircling topological effect also exists in our metasurface design, interestingly in the case of an extremely thin nanostructure. The real and imaginary parts of the eigenvalues obtained by the M-matrix (Eq. (10)) are plotted in Figs. 6(a) and 6(b), respectively, as a function of the complex excitation frequencies. The EP is represented by a yellow star in Fig. 6. It consists of the singular point where the eigenvalue Riemann surfaces cross and coalesce while the imaginary part of the excitation frequency is equal to zero. The black arrows in Fig. 6 reveal that when the EP is encircled, either left- or right-handed, the eigenvalues are interchanged, clearly demonstrating the square root singularity nature of the exceptional point topological vortex that can leading to various innovative properties [56]. Notable, if the spherical nanoparticles of the current metasurface design are changed to chiral active nanostructures with Pancharatnam-Berry or other relevant orientations [7–10], advanced polarization control will be achieved with the advantage of perfect transmission efficiency.

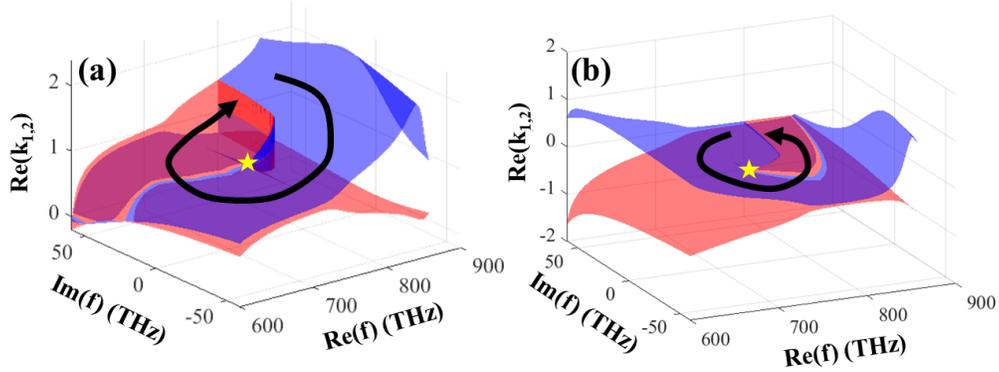

Fig. 6 (a) Real and (b) imaginary parts of the eigenvalues as a function of the complex excitation frequency. The EP is marked with a star at the center of the resulted topological vortex.

The geometric and material parameters of the presented PT-symmetric metasurface can impact its behavior, however without altering the EP formation dynamics. To prove this point, the response of the proposed system as a function of the period $d$, host material permittivity $\varepsilon_h$, and the ratio of the core to shell radii $R = r_s/r_c$ is demonstrated in Fig. 7. More specifically, Figs. 7(a)-(c) show the shift in the resonant wavelength of the passive metasurface, i.e., without gain ($\varepsilon_i = 0$), corresponding to the maximum transmittance point, when the various geometrical parameters vary. As the period $d$ increases, the resonant wavelength slightly decreases. The small degree in the wavelength shift indicates that the resonance response is more dependent on the size of the core and shell coating layer of the nanoparticle instead of the metasurface periodicity. In the case of Fig. 7(b), the shell radius $r_s$ is kept constant to 35nm and the core radius $r_c$ is changed to test different ratios of the core to shell radii ($R = r_s/r_c$). The resonant wavelength of the metasurface strongly depends on the core-shell size ratio and blue shifts as the core radius is decreased compared to the shell radius. Finally, increasing the dielectric constant of the host medium shifts the resonance to higher wavelengths always in the visible range, as demonstrated in Fig. 7(c).

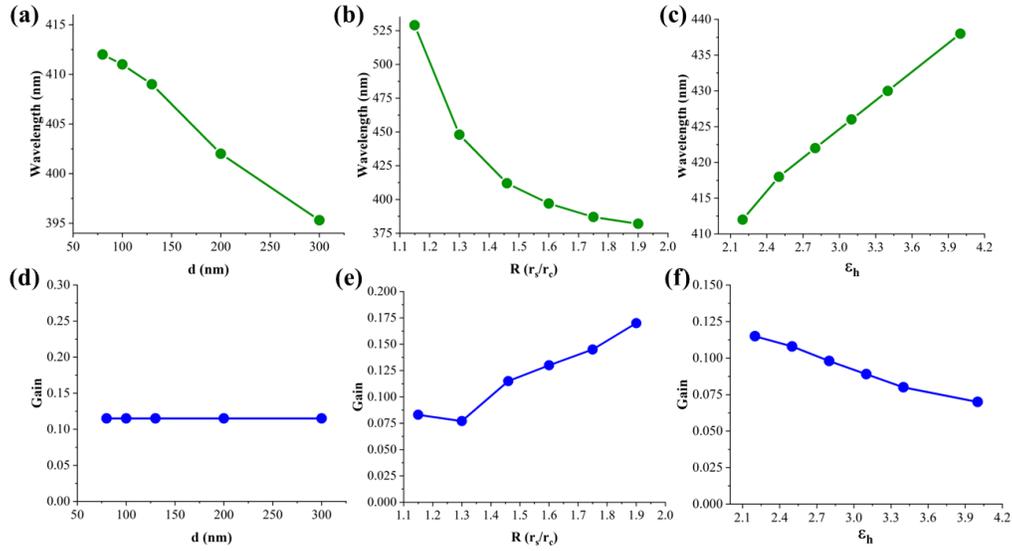

Fig. 7(a)-(c) Resonant transmission wavelength shift of the passive metasurface (no gain) as a function of period $d$, the ratio of the shell to the core radii, and host material permittivity $\varepsilon_h$. (d)-(f) Gain values required to achieve EP formation as a function of $d$, the core-shell radii ratio, and $\varepsilon_h$.

As mentioned before, geometrical variations will not preclude the EP formation. This is shown in Figs. 7(d)-(f) where the gain values required to achieve the EP formation are demonstrated as we vary the metasurface geometric parameters. Interestingly, the gain required to achieve the EP formation does not change with the separation distance, as depicted in Fig. 7(d). Since interactions between the particles become weaker at greater distances, this result indicates that the EP formation is not dependent on the mutual coupling between neighboring nanoparticles and is mainly due to local features of each isolated spherical resonator. Moreover, the gain values required to achieve the EP formation increase with the core-shell radii ratio, as presented in Fig. 7(e). This effect is due to higher plasmonic losses present at lower wavelengths mainly because of the transmission resonance blueshift in the passive system response shown in Fig. 7(b). Finally, the gain values required to achieve the EP formation slightly decrease as the host medium dielectric constant increases, a feature depicted in Fig. 7(f). However, in all aforementioned studies the gain values are always kept low and in experimentally feasible ranges.

Finally, it is worth mentioning that the emergence of the EP is possible to be achieved not only with different geometrical parameters, as demonstrated before in Fig. 7, but also with different material parameters, such as distinct values of permittivity in the dielectric core or alternative plasmonic coating materials (i.e., gold). Furthermore, higher losses in the metallic shell, arising for example from surface dispersion of low thickness silver films [57], do not preclude the EP formation and simply require higher gain to compensate. Hence, the design of the presented non-Hermitian active plasmonic metasurface can be generalized and adapted to a plethora of materials and geometrical parameters without altering its underpinning unique PT-symmetric features in the nanoscale. The analytical results presented in this work have been verified with full-wave simulations and they match to some extent, as was predicted before in relevant works of passive dielectric metasurfaces [43,52]. The obtained analytical results are fast to be obtained and relatively accurate, thus, allowing to achieve an initial design that can be straightforwardly optimized through a more elaborate study based on full-wave simulations, which will be subject of our future work. To sum up, the reflectionless transmission obtained at the EP will enable the use of the proposed active core-shell array metasurface to be employed as a compact nanophotonic optical filter with perfect transmission. Compared to other conventional multilayer optical filter designs [36–39], the proposed metasurface is extremely thin, making its fabrication and implementation much simpler and cost effective.

## 4. Conclusions

To conclude, we have theoretically demonstrated reflectionless transmission in a new PT-symmetric non-Hermitian Huygens' plasmonic metasurface design. The presented active metasurface system exhibits unique characteristics under relatively low gain levels which are always limited to realistic values that can be experimentally verified. Notably, we demonstrate the existence of an EP at a particular wavelength and gain coefficient values, where the eigenvalues of the transfer matrix merge and coalesce in a singular point resulting in reflectionless perfect transmission. We have shown how this EP formation corresponds to the Huygens' dipole response of an individual passive plasmonic composite nanoparticle scatterer. Note that unidirectional transmission is a general phenomenon common to both exceptional points and Huygens' dipoles with the difference that in the exceptional point case the transmission is unitary while metasurfaces based on Huygens' dipoles have high transmission but always lower than one due to losses. Finally, the presented concept of PT-symmetric Huygens' plasmonic metasurface designs is general and is expected to also work for higher order exceptional points obtained by multilayer periodic nanoparticle-based metasurfaces, which will be subject of our future work. The proposed metasurface design is expected to have a plethora of applications in various emerging concepts relevant to nanophotonics, such as in the design of compact optical filters with high transmission efficiency. More generally, the unique connection between PT-symmetry and Huygen's' metasurface physics proposed in the current work is expected to lead to a novel class of optical functionalities in the nascent field of nanophotonics.

**Funding.** This work was partially supported by the Office of Naval Research Young Investigator Program (ONR-YIP) (Grant No. N00014-19-1-2384), the National Science Foundation/EPSCoR RII Track-1: Emergent Quantum Materials and Technologies (EQUATE) (Grant No. OIA-2044049), and the NASA Nebraska Space Grant Fellowship.

**Supplemental document.** See Supplement for supporting content.

# Supplementary Information

# Exceptional points in parity-time symmetric plasmonic Huygens' metasurfaces

Andrew Butler[1] and Christos Argyropoulos[1,2,*]

[1]Department of Electrical and Computer Engineering, University of Nebraska-Lincoln, Lincoln, Nebraska 68588, USA

[2]Department of Electrical Engineering, The Pennsylvania State University, University Park, PA 16803, USA

*christos.argyropoulos@unl.edu

## Evolution of Eigenvalues

In this supplementary section we provide some additional discussion of the evolution of the eigenvalues and their physical effects in the system. To begin, we first investigate what happens to the system with much higher gain compared to the exceptional point response. One interesting behavior occurs at $\varepsilon_i = 1.35$. At this high level of gain, we see the reflectance and transmittance of the system approaching infinity, corresponding to a lasing phenomenon [1,2]. Figure S1 shows the computed reflectance and transmittance spectra as well as the eigenvalues for this level of gain.

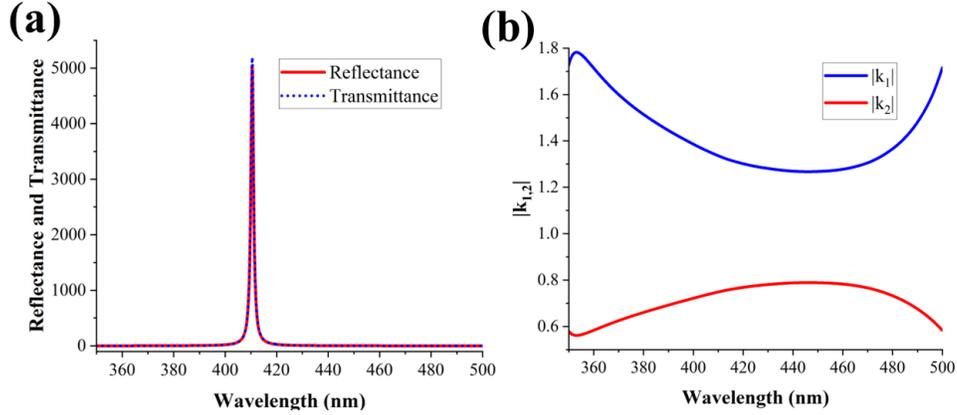

Fig S1. (a) Reflectance and transmittance spectra at the lasing point ($\varepsilon_i = 1.35$). (b) Magnitudes of the eigenvalues for the lasing point as a function of wavelength.

We also explore the evolution of the eigenvalues for higher wavelengths. Figure S2(a) shows the evolution of the eigenvalues of the metasurface over a broader wavelength range compared to Fig. 5(a) in the main paper. Here we use a gain value of $\varepsilon_i = 0.115$, corresponding to the exceptional point formation. At larger wavelengths compared to the exceptional point, the bifurcation of the eigenvalues increases, until one eigenvalue becomes very large, while the other takes almost zero values. At this point (~515nm), it can be seen in Fig. S1(b) that the reflectance of the metasurface is high and the transmittance approaches near zero, which is the exact opposite behavior of the exceptional point located at lower wavelengths. This indicates that a coherent perfect absorber-laser point may be achieved at this wavelength [3]. However, much higher gain values combined with simultaneous counterpropagating illuminations need to be applied to demonstrate this effect [3]. Here, our theoretical method is valid only for single direction illumination, which is the simplest experimental verification method. The more complicated counterpropagating double illumination combined with higher gain value effects are indeed very interesting and will be a subject of our future work.

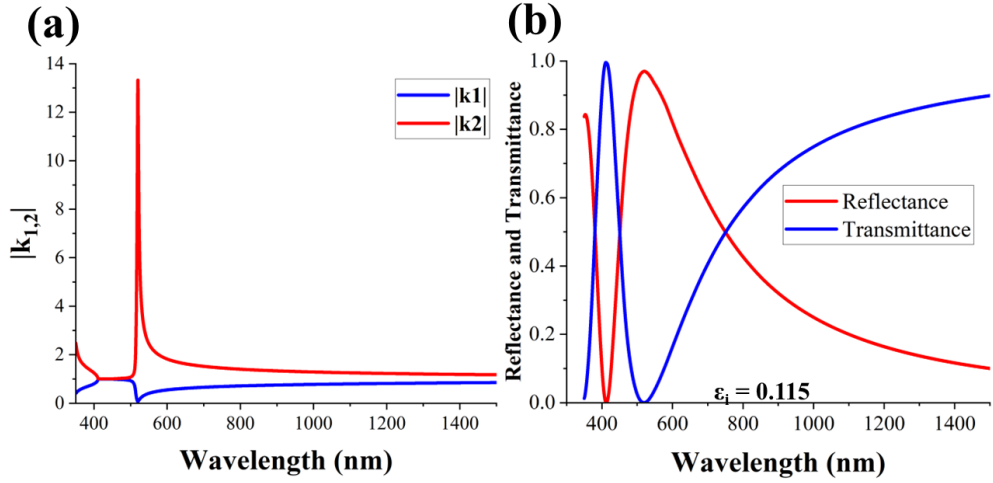

Fig S2. (a) Magnitude of eigenvalues and (b) reflectance and transmittance as a function of wavelength over a broad wavelength range when a gain value of $\varepsilon_i = 0.115$ is used.